\documentclass[seceq]{ptptex}
\usepackage{graphicx}

\pubinfo{Vol.~00, No.~00, May 2007}
\notypesetlogo 
 
\title{Riemann zeta function and quantum chaos}

\author{Eug\`ene \textsc{Bogomolny}}

\inst{CNRS, Universit\'e Paris-Sud, UMR 8626,\\
Laboratoire de Physique Th\'eorique et Mod\`eles Statistiques,\\
91405 Orsay, France}

\recdate{Decemer, 2006}

\abst{A brief review of recent developments in the theory of the Riemann zeta function inspired  by ideas and methods of quantum chaos is given.}

\begin{document}

\maketitle

\section{Introduction}

 At the first glance the Riemann zeta function and quantum chaos are completely disjoint fields. The Riemann zeta function is a part of pure number theory but quantum chaos is a branch of theoretical physics devoted to the investigation of non-integrable quantum problems like the hydrogen atom in external fields. 

Nevertheless for a long time it was understood that there exist multiple interrelations between these two subjects.\cite{Berry_1986,Berry_Keating_1999}   In Sections~\ref{zeta_functions} and \ref{trace_formula}  the Riemann and the Selberg zeta functions and their trace formulae are informally compared.\cite{Hejhal_zeta}  From the comparison  it appears that in many aspects zeros of the Riemann zeta function resemble eigenvalues of an unknown quantum chaotic Hamiltonian. 

One of the principal tools in quantum chaos is the investigation of statistical properties  of deterministic  energy levels of a given Hamiltonian.  In such approach one stresses not precise  values of  physical  quantities but their statistics  by considering them as different realizations of a statistical  ensemble. According to the BGS conjecture\cite{Bohigas_Giannoni_Schmit} energy levels of chaotic quantum systems have the same statistical properties as eigenvalues of standard random matrix ensembles depended only on the exact symmetries.  In Section~\ref{spectral_statistics} it is argued that is quite natural to conjecture  that  statistical properties of the Riemann zeros  are the same as of  eigenvalues of the gaussian unitary ensemble of random matrices (GUE). This conjecture is very well confirmed by numerics but only partial rigorous results are available.\cite{Montgomery} 

In Section~\ref{correlation_functions}  a semiclassical method which permits, in principle,  to calculate correlation functions is shortly discussed. The main problem here  is to control correlations between periodic orbits with almost the same lengths. In 
Sections~\ref{hardy_littlewood_conjecture} and \ref{two_point_riemann} it is demonstrated how the Hardy-Littlewood conjecture\cite{Hardy_Littlewood} about distribution of near-by primes leads to explicit formula for the two-point correlation function of the Riemann zeros.\cite{Bogomolny_Keating_1996} The resulting formula  describes non-universal approach  to the GUE result in excellent  agreement with numerical results. 

In Section~\ref{nearest_neighbor}  it is demonstrated how  to calculate non-universal corrections to the nearest-neighbor distribution for the Riemann zeros.\cite{nearest_neighbor_2006} 

Spectral statistics is not the only interesting statistical characteristics of zeta functions. The mean moments of the Riemann zeta function along the critical line is another important subject that attracts wide attention in number theory during a long time. In Section~\ref{zeta_moments} it is explained how random matrix theory permit Keating and Snaith to propose the breakthrough  conjecture about mean moments.\cite{Keating_Snaith} This conjecture now is widely accepted and is generalized for different zeta and $L$-functions and different quantities as well.\cite{group,Conrey_Snaith}.
  
\section{Riemann and Selberg zeta functions}\label{zeta_functions}

The    Riemann zeta function, $\zeta(s)$, is defined as  
\begin{equation}
\zeta (s)=\sum_{n=1}^{\infty}\frac{1}{n^s}
\label{zeta}
\end{equation}
where the summation is taken over all integers and $s$ is a complex parameter (see e.g. \cite{Edwards}). As each integer can uniquely (up to the ordering) be written as a product over prime numbers $p$, $n=p_1^{m_1}p_2^{m_2}\ldots p_k^{m_k}$, the sum in (\ref{zeta}) may be expressed  as the Euler product over all prime numbers
\begin{equation}
\zeta(s)=\prod_p \frac{1}{(1-p^{-s})}\;. 
 \label{zeta_product}
 \end{equation}
The Selberg zeta function, $Z(s)$, is related with hyperbolic motion on constant curvature surfaces generated by discrete groups.\cite{Hejhal} Similarly to (\ref{zeta_product}) it is defined as the product but not over primes but over all primitive periodic orbits (ppo) for the motion on the surface considered
\begin{equation}
Z(s)=\prod_{{\rm ppo}}\prod_{m=0}^{\infty}(1-{\rm e}^{-l_p(s+m)})\;.
\label{Selberg}
\end{equation}
Here  $l_p$ are the lengths of these orbits and $s$ is a complex number. (An informal introduction to this subject is given e.g. in Ref.~\citen{LesHouches}.)

\begin{figure}
\begin{minipage}{.48\linewidth}
\centerline{\includegraphics[width=.99\linewidth]{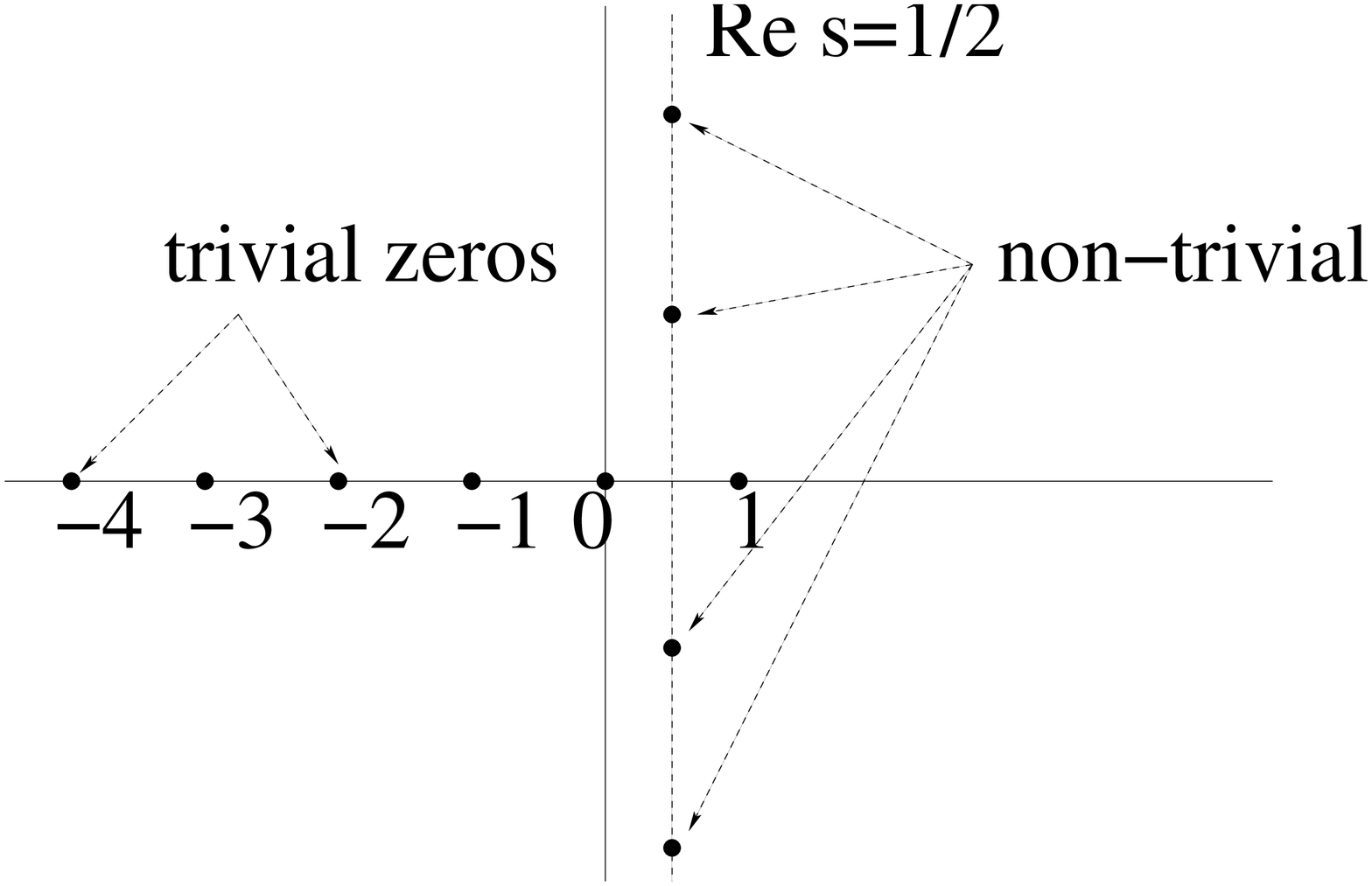}}
\end{minipage}
\begin{minipage}{.52\linewidth}
\centerline{\includegraphics[angle=-90,width=.99\linewidth]{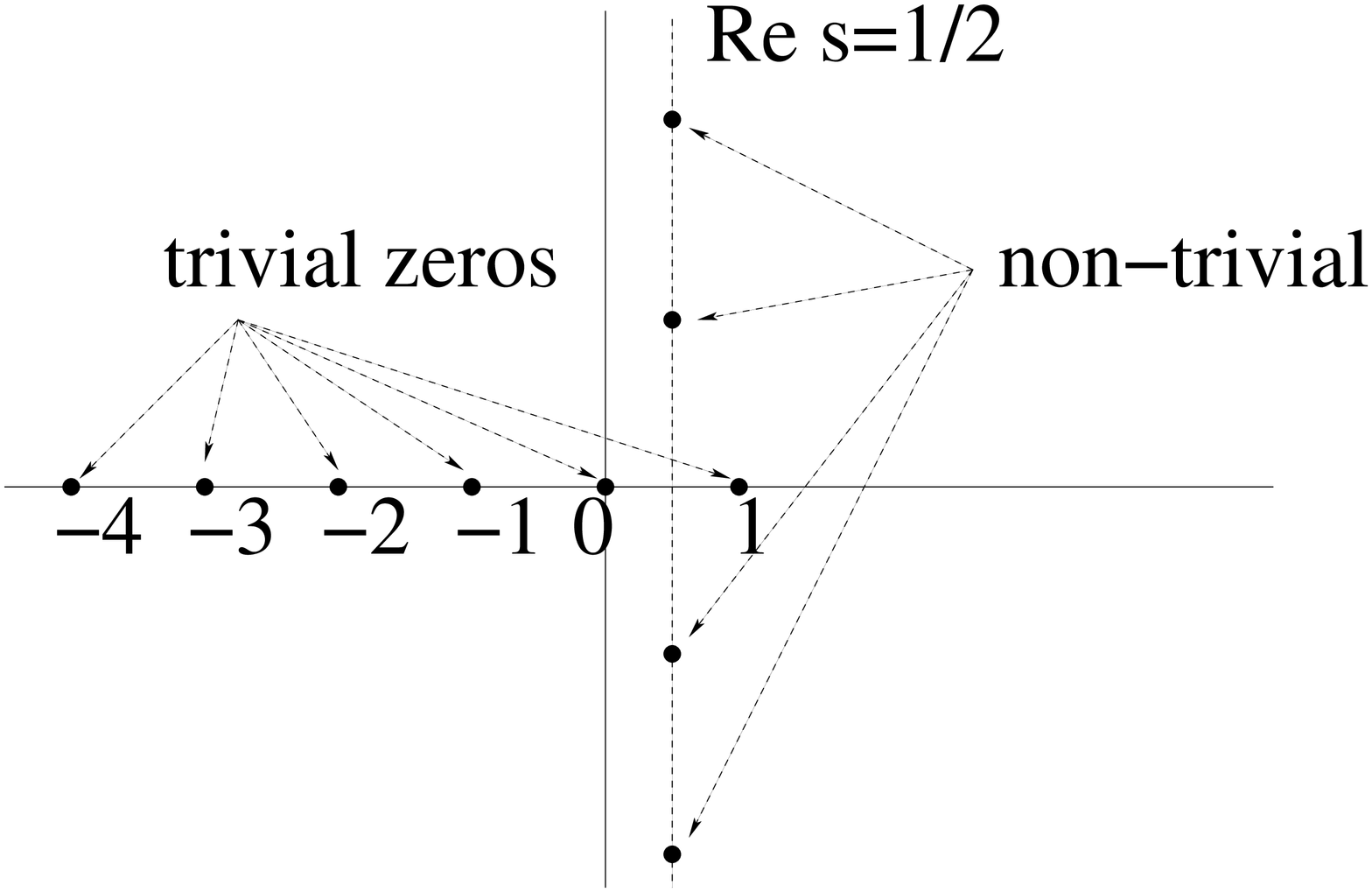}}
\end{minipage}
\caption{Structure of zeros for the Riemann  (left) and the Selberg (right) zeta functions. The Riemann zeta function has the pole at $s=1$ and trivial zeros are at $s=-2,-4,\ldots, $. Trivial zeros for the Selberg zeta function with compact domain are at $s=1,0,-1,-2,\ldots, $.}
\label{fig1} 
\end{figure}

The  zeta functions (\ref{zeta_product}) and (\ref{Selberg}) converge only when Re$(s)>1$ but the both  can analytically be continued into the whole complex plane of $s$ where they obey similar functional equations
\begin{equation}
\zeta(s)=\phi_R(s)\zeta(1-s)\;,\;\;\;Z(s)=\phi_S(s)Z(1-s)
\label{functional_zeta}
\end{equation}
where for the Riemann zeta function  
\begin{equation}
\phi_R(s)=2^s\pi^{s-1}\sin \left ( \pi s/2 \right ) \Gamma (1-s)\;.
\end{equation}
For the Selberg case 
the function $\phi_S(s)$ depends on the group considered. In the simplest case of groups with compact fundamendal domain with area $\mu$
\begin{equation}
\phi_S(s) =\exp \left (\mu\int_0^{s-1/2}u\tan \pi u\  {\rm d}u\right )\;.
\end{equation}
The both zeta functions have also a striking resemblance of their zero structure (see Fig.~\ref{fig1}).  There exist two types of zeros: trivial ones coming from  known zeros of the factor $\phi(s)$ and non-trivial ones which lie on the line  Re$(s)=1/2$.  For the Selberg zeros this property is a theorem because  its non-trivial zeros are related with eigenvalues of the self-adjoint Laplace--Beltrami operator on constant negative curvature surfaces. For the Riemann zeros it is the content of the famous Riemann conjecture which is confirmed by all existing numerical calculations.  

\section{Trace formulae}\label{trace_formula}

Non-trivial zeros are not known analytically (cf. Ref.~\citen{Odlyzko_home} where lowest Riemann zeros are given with 1000 digits) but for the both zeta functions there exist trace formulae which express a sum over all non-trivial zeros, $s_n$, through a sum over  primes for the Riemann case and over periodic orbits for the Selberg case (see e.g. Ref.~\citen{Hejhal_zeta}).

Let $h(r)$ be a test function with the following properties 
\begin{itemize}
\item $h(r)$ is analytical in $|\mbox{Im }r|\leq 1/2+\delta$,
\item $h(-r)=h(r)$,
\item $|h(r)|\leq A(1+|r|)^{-2-\delta}$,
\end{itemize}  
and  $g(u)$ be the Fourier transform of $h(r)$
$$
g(u)=\frac{1}{2\pi} \int_{-\infty}^{\infty} h(r){\rm e}^{-{\rm i}ru}{\rm d}r\;.
$$
The trace formula for  Riemann zeros (Weil explicit formula) is 
\begin{eqnarray}
& &\sum_{\begin{array}{c}\mbox{\scriptsize{non-trivial}}\\\mbox{ \scriptsize{zeros}}
\end{array}}h( s_n)
=\frac{1}{2\pi}\int_{-\infty}^{\infty}
h(r)\frac{\Gamma'}{\Gamma}(\frac{1}{4}+\frac{i}{2}r){\rm d}r-\label{trace_riemann}\\
&-&2\sum_{\mbox{\scriptsize{primes}}}\ln p\sum_{n=1}^{\infty}
  \frac{1}{{\rm e}^{n\ln p/2}}g(n\ln p)+ h(\frac{{\rm i}}{2})+h(-\frac{{\rm i}}{2})-g(0)\ln \pi
\nonumber
\end{eqnarray}
The trace formula for  zeros of the Selberg zeta function (for groups with finite fundamental domains) has similar form
\begin{equation}
\sum_{\begin{array}{c}\mbox{\scriptsize{non-trivial}}\\\mbox{\scriptsize{ zeros}}
\end{array}}h(s_n)
=\frac{\mu}{2\pi}\int_{-\infty}^{\infty}
h(r)r\tanh (\pi r){\rm d}r+
\sum_{\mbox{\scriptsize{ppo}}}l_p\sum_{n=1}^{\infty}\frac{1}{2\sinh(nl_p/2)}g(nl_p)\;.
\label{trace_selberg}
\end{equation}
Here  $l_p$ are length of periodic orbits and $\mu$ is the hyperbolic area of the fundamental domain.

 The Selberg trace formula is exact. For general chaotic systems Gutzwiller\cite{Gutzwiller_1,Gutzwiller_2}  obtained semiclassical  trace formula which, in general,  is only the first dominant term in the semiclassical limit $\hbar\to 0$ (though the calculation of correction terms  are in principle possible\cite{Gaspard}). For a two dimensional chaotic billiard with area $\mu$  the Gutzwiller trace formula for the density of states in 'semiclassical' limit $E\to \infty$ is   the sum of two terms
\begin{equation}
d(E)\equiv \sum_n\delta(E-E_n)=\bar{d}(E)+d^{(osc)}(E)
\label{density_states}
\end{equation}
 where the smooth part  is given by the usual Thomas--Fermi expression:
$\bar{d}(E)=\mu/4\pi$,
and the oscillatory contribution is represented as the sum over all classical periodic orbits with coefficients obtained from purely classical mechanics  
\begin{equation}
d^{(osc)}(E)=\sum_{\mbox{\scriptsize ppo}}\frac{T_p }{ \hbar}\sum_{n=1}^{\infty}
\frac{1}{|\det (M_p^n-1)|^{1/2}}\cos \left (n(k l_p-\frac{\pi}{2}\mu_p)\right ).
\end{equation}
Here $k=\sqrt{E}$ is the momentum and for each primitive periodic orbit, $p$, $l_p$ is its length, $M_p$ is the monodromy matrix, and $\mu_p$ is the Maslov index.\cite{Gutzwiller_2} 

Ignoring questions of convergence, trace formulae for the Riemann and the Selberg zeta functions,  (\ref{trace_riemann}) and (\ref{trace_selberg}), can  also be rewritten in similar form by taking  $h(r)=\delta(r-E)$ and conserving only terms dominant at large energy. In particular for the Riemann zeros such 'physical'  trace formula takes the form (\ref{density_states})  with
\begin{equation}
\bar{d}(E)=\frac{1}{2\pi}\ln\frac{E}{2\pi}
\label{mean_density}
\end{equation}
and
\begin{equation}
d^{(osc)}(E)=-\frac{1}{\pi}\sum_p\sum_{n=1}^{\infty} \frac{\ln p}{p^{n/2}}\cos(En\ln p)
=-\frac{1}{\pi}\sum_{n=1}^{\infty} \frac{1}{\sqrt{n}}
\Lambda(n)\cos(E\ln n)\;.
\label{trace_lambda}
\end{equation}
where $\Lambda(n)$ is von Mangoldt function
\begin{equation}
\Lambda(n)=\left \{ \begin{array}{lr}
\ln p, & \mbox{if } n=p^k\\
0, & \mbox{otherwise}
\end{array} .\right .
\label{lambda}
\end{equation}
The above formulae demonstrate a striking resemblance between  the Riemann and the Selberg  zeta functions.   Different quantities for number-theoretical zeta functions   have analogs in dynamical zeta functions (and vise verso) according to  the following dictionary\cite{Berry_1986,Hejhal_zeta}

\vspace{.2in}

\fbox{\begin{minipage}[t]{.45\linewidth}
 \begin{center} {\large Riemann zeta function}\end{center}
 
 \vspace{.5cm}

\begin{flushright}

{ Primes} $\longrightarrow$

 $\ln p$ $\longrightarrow$
 
 non-trivial zeros $\longrightarrow$
 \end{flushright}

\vspace{.5cm}

\begin{center} Prime number theorem: \end{center}

\begin{flushright}
$$
N(\ln p<T)\stackrel{T\to \infty}{\longrightarrow}\;\frac{\exp(T)}{T}\;\;\;\longrightarrow
$$

\vspace{.5cm}

\end{flushright}
\end{minipage}}
\fbox{\begin{minipage}[t]{.45\linewidth}
\begin{center} {\large  Selberg zeta function}\end{center}

\vspace{.5cm}

\begin{flushleft}

{ $\longleftarrow$ Periodic orbits} 

$\longleftarrow$ {Period} $T_p$

$\longleftarrow$ eigen momenta  
\end{flushleft}

\vspace{.5cm}

\begin{center}Periodic orbit number theorem:\end{center}

\begin{flushleft}
$$
\longleftarrow \;\;\;\;N(T_p<T)\stackrel{T\to \infty}{\longrightarrow}\;\frac{\exp(T)}{T}
$$

\vspace{.5cm}

\end{flushleft}
\end{minipage}}

\vspace{.2in}

It opens an exiting possibility (cf. Refs.~\citen{Berry_1986,Hejhal_zeta}) that the Riemann zeta function is similar to a dynamical zeta function for a chaotic quantum system corresponding to the motion on a constant negative curvature surface.  Till now all  attempts to find such system failed (cf. Refs.~\citen{Connes} and \citen{xp}) but a quete of a self-adjoint operator whose eigenvalues are related with non-trivial Riemann zeros (as it seems to be suggested by Polya and Hilbert) continue. Nevertheless notice that the overall signs of the trace formulas for the Riemann zeta and dynamical systems are opposite which complicates such interpretation (cf. in this connection Connes' absorbtion spectrum\cite{Connes}).
  
\section{Spectral statistics}\label{spectral_statistics}

If one accepts that the Riemann zeta function is analogous to a dynamical zeta function and its zeros are in a certain sense related to eigenvalues of an unknown Hamiltonian it is tempting to applied to Riemann zeros numerous results derived within dynamical quantum chaos. 

First of all, it is natural to investigate  statistical  properties  of Riemann zeros. Of course, Riemann zeros are defined by a deterministic procedure and are not random. Nevertheless, if one takes groups of $N$ consecutive  zeros in different part of the spectra such collection as a whole can be considered as originated from realizations of a random ensemble.     

According to the Bohigas, Giannoni, Schmit conjecture\cite{Bohigas_Giannoni_Schmit}  energy levels of chaotic quantum systems on the scale of the mean level density are distributed as  eigenvalues of standard  random matrix ensembles depended  only on symmetry properties. There are three main classes of universality\cite{Mehta}. Integer spin systems without (resp. with) time-reversal invariance are described by  Gaussian Unitary Ensemble (GUE) (resp. Gaussian Orthogonal Ensemble (GOE)) and systems with half-integer spin and time-reversal invariance are supposed to belong to Gaussian Symplectic Ensemble (GSE). 

Symmetry properties manifest primarily in the degeneracies of periodic orbit lengths. Time-reversal invariant systems should have doublets of periodic orbits with exactly the same length, but systems without time-reversal invariance, in general, have no exact degeneracies. In the above analogy prime numbers play the role of periodic orbits. As primes are not degenerated the conjectural dynamical system connected with the Riemann zeros should belong to the universality class of time-non-invariant systems and, consequently, it should have spectral statistics as gaussian unitary ensemble of random matrices (GUE).     

It is well known \cite{Mehta} that the joint distribution of eigenvalues, $E_j$, for this ensemble has the following form  
\begin{equation}
P(E_1,E_2,\ldots,E_N)\sim \prod_{i<j}|E_i-E_j|^{2}\exp (-\sum_{k=1}^N E_k^2)\;.
\label{GUE}
\end{equation}
The knowledge of the joint distribution of eigenvalues permits formally to calculate $n$-point correlation functions which are defined as the probability density  of having $n$ eigenvalues at given positions
\begin{equation}
R_n(E_1,\ldots,E_n)\sim \int P(E_1,\ldots,E_N)dE_{n+1}\ldots {\rm d}E_{N}\;.
\end{equation}
For classical random matrix ensembles correlation functions are know analytically \cite{Mehta}. For GUE (\ref{GUE}) they have the determinantal form 
\begin{equation}
R_n(E_1,\ldots,E_n)=\det [ K(E_i,E_j)]|_{i,j=1,\ldots,n}
\label{R_n}
\end{equation}
where in the universal limit of unit mean density and $N\to \infty$ the kernel $K(E_i,E_j)$ is 
\begin{equation}
K(E_i,E_j)=\frac{\sin \pi(E_i-E_j)}{\pi (E_i-E_j)}\;.
\label{kernel}
\end{equation}
In particular, the unfolded two-point correlation function has especially simple form
\begin{equation}
R_2(\varepsilon)=1-\left (\frac{\sin \pi \varepsilon}{\pi \varepsilon}\right )^2=1-\frac{1}{2\pi^2\varepsilon^2}+\frac{1}{4\pi^2\varepsilon^2}({\rm e}^{2 {\rm i}\varepsilon}+{\rm e}^{-2 {\rm i} \varepsilon})
\label{R_2}
\end{equation}
where $\varepsilon=E_1-E_2$.

The above discussion leads to the conjecture that correlation functions for high Riemann zeros are given by Eqs.~(\ref{R_n})-(\ref{R_2}). It was Montgomery\cite{Montgomery} who in 1973 first proposed this conjecture. He considered two-point correlation function of Riemann zeros and proved the result equivalent to the existence of $1/(2\pi^2\varepsilon^2)$ term in (\ref{R_2}). Physically Montgomery's theorem  corresponds to the diagonal approximation (see below).  Though oscillating  terms in  (\ref{R_2}) cannot be obtained rigorously, Montgomery conjectured that spectral statistics of  Riemann zeros coincides with GUE statistics.
  
\begin{figure}
\begin{minipage}{.45\linewidth}
\begin{center}
\includegraphics[angle=-90, width=.99\linewidth,  bb=120 0 550 640,clip=]{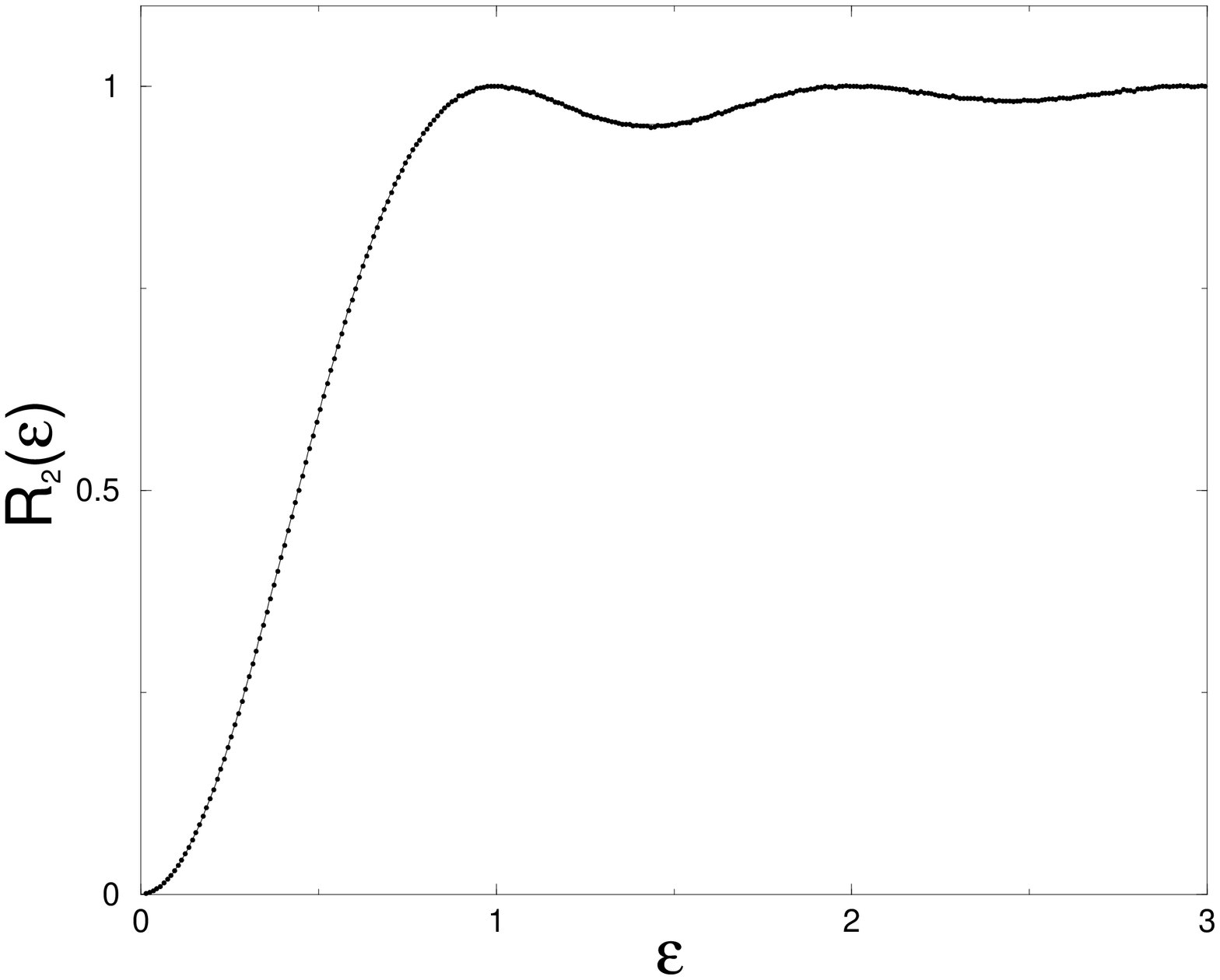}
\end{center}
\end{minipage}
\begin{minipage}{.45\linewidth}
\begin{center}
\includegraphics[width=.99\linewidth, bb=0 0 290 211, clip=]{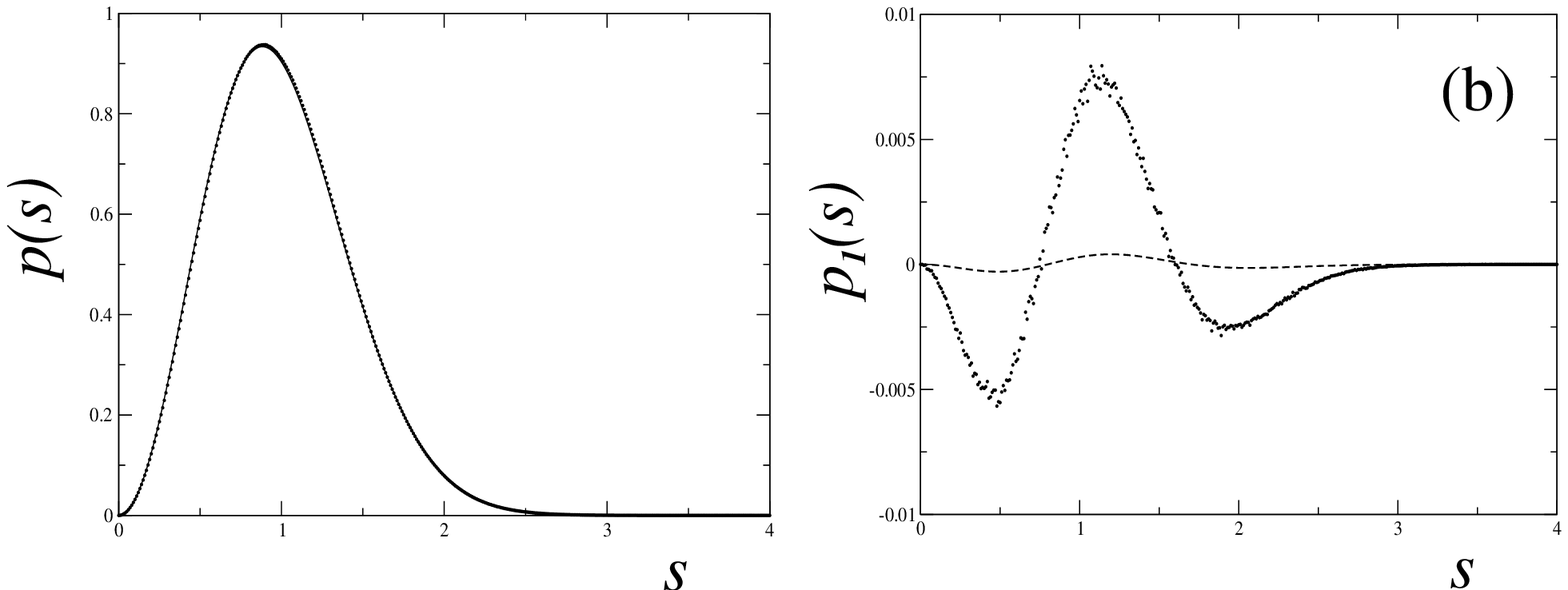}
\end{center}
\end{minipage}
\caption{Left: Two-point function for $2\cdot 10^{8}$ Riemann zeros near the $10^{23}$-th zero. Solid line: GUE prediction (\ref{R_2}). Right: Nearest-neighbor distribution for $10^9$ Riemann zeros close to $E=2.5\cdot 10^{15}$. Solid line: GUE prediction.}
\label{fig2}
\end{figure}

Odlyzko has performed very large scale computation of Riemann zeros\cite{Odlyzko}. In Fig.~\ref{fig2} left the  two-point correlation function numerically computed by  Odlyzko  is presented. It is clearly seen that the agreement between numerical calculations and GUE prediction (\ref{R_2})  is excellent.
 Another important statistics of Riemann zeros, namely, the distribution of the nearest neighbor is shown in Fig.~\ref{fig2} right. Once more the agreement with GUE is striking. 

\section{Correlation functions}\label{correlation_functions}

In view of such excellent confirmation of GUE conjecture for Riemann zeros it is natural to try to derive (at least physically) correlation functions from the first principles.
  
Formally the $n$-point correlation function is defined as the
probability of having $n$ energy levels at given positions.
\begin{equation}
R_n(\epsilon_1,\ldots,\epsilon_n)=
\left <d(E+\epsilon_1)d(E+\epsilon_2)\ldots d(E+\epsilon_n)\right >\;.
\end{equation}
Here $\left <\dots\right >$ denotes an energy smoothing 
\begin{equation}
\left <f(E)\right >=\int f(E')\sigma (E-E'){\rm d}E'\; .
\label{sigma}
\end{equation}
$\sigma(E)$ is a smoothing function picked at zero (like the Gaussian). It  is assumed to be normalized  by the condition  $\int \sigma(E){\rm d}E=1$ and to have a certain small width  $\Delta E$.  

The simplest way of the using these formulae is to substitute instead of the true density $d(E)$ its semiclassical expression 
\begin{equation}
d(E)=\bar{d}(E)+\sum_{p,n}A_{p,n}{\rm e}^{{\rm i} n S_p(E)/\hbar} +{\rm c.c.}\;.
\end{equation}
It gives the following explicit formula for the two-point correlation function
\begin{equation}
R_2(\epsilon_1, \epsilon_2)=\bar{d}^2+\sum_{p_i,n_i} A_{p_1,n_1}A_{p_2,n_2}^*
\left < \exp \frac{{\rm i}}{\hbar}\left [n_1S_{p_1}(E+\epsilon_1)-n_2S_{p_2}(E+\epsilon_2)\right ]\right > +{\rm c.c.}
\end{equation}
where
$S_p(E+\epsilon)=S_p+T_p\epsilon$  and  $T(E)$ is the classical period of motion. 
Finally one gets
\begin{equation}
R_2(\epsilon_1, \epsilon_2)=\bar{d}^2+\sum_{p_i,n_i} A_{p_1,n_1}A_{p_2,n_2}^*  B(p_i,n_i)
\exp \frac{{\rm i}}{\hbar}\left [n_1 T_{p_1}(E)\epsilon_1-n_2 T_{p_2}(E)\epsilon_2\right ]+{\rm c.c.}
\end{equation}
and 
\begin{equation}
B(p_i, n_i)=\left <\exp \frac{{\rm i}}{\hbar}\left [n_1 S_{p_1}(E)-n_2 S_{p_2}(E)\right ] \right >\;.
\end{equation}
The calculation of this quantity is the main problem in the semiclassical approach to spectral statistics. To get the answer correctly one has to take into account tiny correlations between lengths of different periodic orbits which is in general a quite non-trivial task.\cite{Varenna}   

\section{Hardy-Littlewood conjecture}\label{hardy_littlewood_conjecture}
  
Using the trace formula for the Riemann zeta function  (\ref{trace_lambda}) one gets that 
the connected two-point correlation function of Riemann zeros $R_2^{(c)}=R_2-\bar{d}^2$
\begin{equation}
R_2^{(c)}(\epsilon_1,\epsilon_2)=\frac{1}{4\pi^2}\sum_{n_1,n_2}
\frac{\Lambda(n_1)\Lambda(n_2)}{\sqrt{n_1n_2}}\\
\left <{\rm e}^{{\rm i}(E+\epsilon_1)\ln n_1-{\rm i}(E+\epsilon_2)\ln n_2}\right > +{\rm c.c.}\;.
\end{equation}
Here $\left <\ldots\right >$ denotes an averaging over $E$ as in (\ref{sigma}). The simplest approximation (called the diagonal approximation\cite{Berry_diagonal}) consists in taking into account only pairs of terms with exactly the same periodic orbit length (i.e. $n_1=n_2$).  Then 
\begin{equation}
R_2^{(diag)}(\epsilon_1,\epsilon_2)=\frac{1}{4\pi^2}\sum_{n}
\frac{\Lambda^2(n)}{n}
{\rm e}^{{\rm i}(\epsilon_1-\epsilon_2)\ln n} +{\rm c.c.}\;.
\end{equation}
By straightforward calculations this expression may  be transformed as follows ($\epsilon=\epsilon_1-\epsilon_2$)
\begin{equation}
R_2^{(diag)}(\epsilon)=-\frac{1}{4\pi^2}
\frac{\partial^2}{\partial \epsilon ^2}\ln\left ( |\zeta(1+{\rm i}\epsilon)|^2\Phi^{(diag)}(\epsilon)\right )
\label{riemann_diag}
\end{equation}
and $\Phi^{(diag)}(\epsilon)$ is given by the convergent sum over primes 
\begin{equation}
\Phi^{(diag)}(\epsilon)=\exp (\sum_p\sum_{m=1}^{\infty} 
\frac{1-m}{m^2p^m}{\rm e}^{{\rm i}m\ln p \epsilon}+{\rm c.c.})\;.
\end{equation}
In the limit $\epsilon \rightarrow 0$, 
$\zeta(1+{\rm i}\epsilon)\rightarrow ({\rm i}\epsilon)^{-1}$ and
$\Phi^{(diag)}(\epsilon)\rightarrow$ const. Hence
$$
R_2^{(diag)}(\epsilon)\rightarrow -\frac{1}{2\pi^2\epsilon^2},
$$
which agrees with the smooth part of the GUE result (\ref{R_2}) and corresponds to Montgomery's theorem \cite{Montgomery} .

The calculation of off-diagonal contribution is more difficult. Formally it  can be expressed   as the double sum over primes
\begin{equation} 
R_2^{(off)}(\epsilon_1,\epsilon_2)=\frac{1}{4\pi^2}\sum_{n_1\neq n_2}
\frac{\Lambda(n_1)\Lambda(n_2)}{\sqrt{n_1n_2}}
\left <{\rm e}^{{\rm i}E\ln (n_1/n_2)+{\rm i}(\epsilon_1 \ln n_1-\epsilon_2\ln n_2)}\right >+{\rm c.c.}\; .
\end{equation}
The exponent $\exp ({\rm i}E\ln (n_1/n_2))$ oscillates quickly except  for $n_1$ close to  $n_2$.
One writes $n_1=n_2+r$ and  $\ln (n_1/n_2)\approx r/n_2$. In this manner $R_2^{(off)}(\epsilon_1,\epsilon_2)$ is rewritten as follows 
\begin{equation}
R_2^{(off)}(\epsilon)=\frac{1}{4\pi^2}\sum_{n,r}
\frac{\Lambda(n)\Lambda(n+r)}{n}
\left <{\rm e}^{{\rm i}E r/n+{\rm i}\epsilon \ln n}\right > +{\rm c.c.}\;.
\end{equation}
To proceed further one needs to control the behavior of the the product of two von Mangolt functions $\Lambda(n)\Lambda(n+r)$ which is  nonzero only when both $n$ and $n+r$ are power of prime numbers. 

It is naturally to assume that the
dominant contribution to the above sum comes from the mean value of this product
\begin{equation}
\alpha (r)=\lim_{N\rightarrow \infty}\frac{1}{N}\sum_{n=1}^N\Lambda(n)\Lambda(n+r)\;.
\end{equation}
No exact theorems fix the value of this limit. It  is  known only from the famous Hardy-Littlewood conjecture\cite{Hardy_Littlewood} according to which it is expresses through the following singular series
\begin{equation}
\alpha(r)=\sum_{(p,q)=1}{\rm e}^{2\pi {\rm i}pr/q} \left (\frac{\mu(q)}{\varphi(q)}\right )^2\;.
\label{HL}
\end{equation}
Here the sum is taken over all integers $q=1,2,\ldots $ and all integer $p<q$ co-prime to $q$.  $\mu(q)$ is the M\"obius function
\begin{equation}
\mu(q)=\left \{ \begin{array}{cl}1&\mbox{ if }q=1\\(-1)^k&\mbox{ if }q=p_1\ldots p_k\\
0&\mbox{ if } q \mbox{ is divisible on }p^2\end{array}\right .
\end{equation}
and $\varphi(q)$ is the  Euler function which counts integers less than $q$ and coprime to $q$ 
\begin{equation}
\varphi(q)=q\prod_{p|q}\left (1-\frac{1}{q}\right )\;.
\end{equation}
Using properties of such  series one proves\cite{Hardy_Littlewood} that $\alpha(r)$  for even $r$  can be represented  as the finite product 
\begin{equation}
\alpha (r)=C_2\prod_{p|r}\frac{p-1}{p-2}\;.
\end{equation}
The product is taken over all prime divisors of $r$ except $2$  and
$C_2$ is called the  twin prime constant
\begin{equation}
C_2=2\prod_{p>2}\left (1-\frac{1}{(p-1)^2}\right )\approx 1.32032\ldots .
\end{equation}
Physically the Hardy-Littlewood conjecture gives the number of prime pairs such that $p_1-p_2=r$ and $p_i<N$  
\begin{equation}
N(p,p+r \mbox{ are primes and }p<N)\stackrel{N\to\infty}{\longrightarrow}\frac{N}{\ln^2 N}\alpha(r)\;.
\end{equation}
In particular it predicts that the number of twins primes (i.e.  $p_1-p_2=2$) is asymptotically  $C_2N/\ln^2 N$. This conjecture is well confirmed by existing numerics but no rigorous proof exists. The heuristic derivation of the Hardy-Littlewood conjecture is given e.g. in Ref.~\citen{LesHouches}. 
  
\section{Two-point correlation function of Riemann zeros}\label{two_point_riemann}

Taking  the above formulae as granted one gets
\begin{equation}
R_2^{(off)}(\epsilon)=\frac{1}{4\pi^2}\sum_n
\frac{1}{n}{\rm e}^{{\rm i}\epsilon \ln n}
\sum_r \alpha(r){\rm e}^{{\rm i}E r/n} +{\rm c.c.}\;.
\end{equation}
After substitution the Hardy-Littlewood formula for $\alpha(r)$  (\ref{HL}) and performing the sum over all $r$ one obtains
\begin{equation}
R_2^{(off)}(\epsilon)=\frac{1}{4\pi^2}\sum_n
\frac{1}{n}{\rm e}^{{\rm i}\epsilon \ln n}
\sum_{(p,q)=1}\left (\frac{\mu(q)}{\psi(q)}\right )^2
\delta(\frac{p}{q}-\frac{E}{2\pi n}) +{\rm c.c.}\;.
\end{equation}
Changing the sum over $n$ to the integral gives
\begin{equation}
R_2^{(off)}(\epsilon)=\frac{1}{4\pi^2}{\rm e}^{{\rm i}\epsilon \ln \frac{E}{2\pi}}
\sum_{(p,q)=1}\left (\frac{\mu(q)}{\psi(q)}\right )^2
(\frac{q}{p})^{1+{\rm i}\epsilon} +{\rm c.c.}\;.
\end{equation}
The sum over coprime integers can be performed by using the inclusion-exclusion principle 
\begin{equation}
\sum_{(p,q)=1}f(p)=\sum_{k=1}^{\infty}\sum_{\delta |q}f(k\delta)\mu(\delta),
\end{equation}
and taking into account that $2\pi \bar{d}=\ln (E/2\pi)$. Finally  one obtains\cite{Bogomolny_Keating_1996} 
\begin{equation}
R_2^{(off)}(\epsilon)= \frac{1}{4\pi^2}|\zeta(1+{\rm i}\epsilon)|^2
{\rm e}^{2\pi {\rm i}\bar{d}\epsilon }\Phi^{(off)}(\epsilon)+{\rm c.c.}\;.
\label{riemann_off}
\end{equation}
The function $\Phi^{(off)}(\epsilon)$ is given by a  convergent product over all primes
\begin{equation}
\Phi^{(off)}(\epsilon)= \prod_p\left (1-\frac{(1-p^{{\rm i}\epsilon})^2}{(p-1)^2}\right )\;.
\end{equation}
From the above formulae  the two-point  correlation function of  Riemann zeros  is the sum of three terms  
\begin{equation}
R_2(\epsilon)=\bar{d}^2(E)+R_2^{(diag)}(\epsilon)+R_2^{(off)}(\epsilon)
\label{exact_R}
\end{equation}
where the mean density of the Riemann zeros, $\bar{d}(E)$, is given in (\ref{density_states}), the diagonal, $R_2^{(diag)}(\epsilon)$, and off-diagonal, $R_2^{(off)}(\epsilon)$, contributions are defined in (\ref{riemann_diag}) and (\ref{riemann_off}) respectively. 

The very optimistic error in these formulae is supposed to be of the order of $1/\sqrt{E}$ which means that for all  practical reasons they can be considered as the exact ones. 

To compare them with numerics it is necessary first to perform the unfolding of the spectrum. It is done by measuring the distance between zeros in units of the mean zeros density and re-scaling  the correlation function as follows
\begin{equation}
R_2(\varepsilon)=\frac{1}{\bar{d}^2(E)}R_2(\frac{\varepsilon}{\bar{d}(E)})\;.
\end{equation}
Here $\varepsilon$ is considered as a finite value. 

The limit $E\to\infty$ implies $\bar{d}(E)\to\infty$ and the above formulae tend to the universal GUE result. 
\begin{equation}
R_2^{(off)}(\varepsilon)\stackrel{E\to\infty}{\longrightarrow} \frac{1}{(2 \pi \varepsilon)^2}
({\rm e}^{2\pi {\rm i}\varepsilon }+{\rm e}^{-2\pi {\rm i}\varepsilon })
\end{equation}
 which agrees  with oscillating terms of the GUE prediction (\ref{R_2}). In Ref.~\citen{Bogomolny_Keating_1995} it was demonstrated that in this limit higher order correlation functions of the Riemann zeros also tend to the GUE results.
  
The importance of the above formulae comes first of all from the fact that they describe not only the universal (structureless) GUE limit but also  non-universal approach to it. 

\begin{figure}
\begin{minipage}{.45\linewidth}
\includegraphics[angle=-90, width=.99\linewidth]{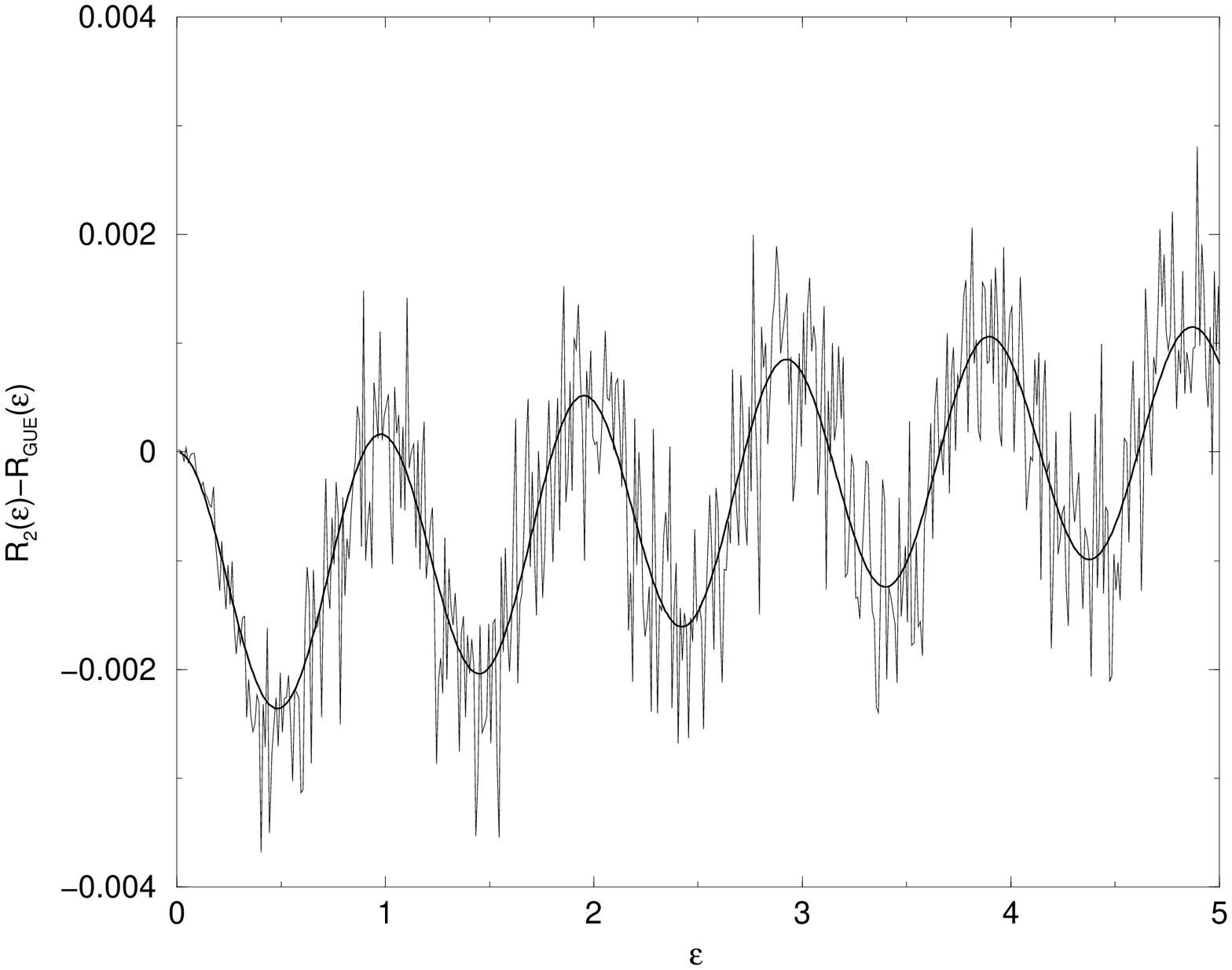}
\end{minipage}
\begin{minipage}{.45\linewidth}
\includegraphics[angle=-90, width=.99\linewidth]{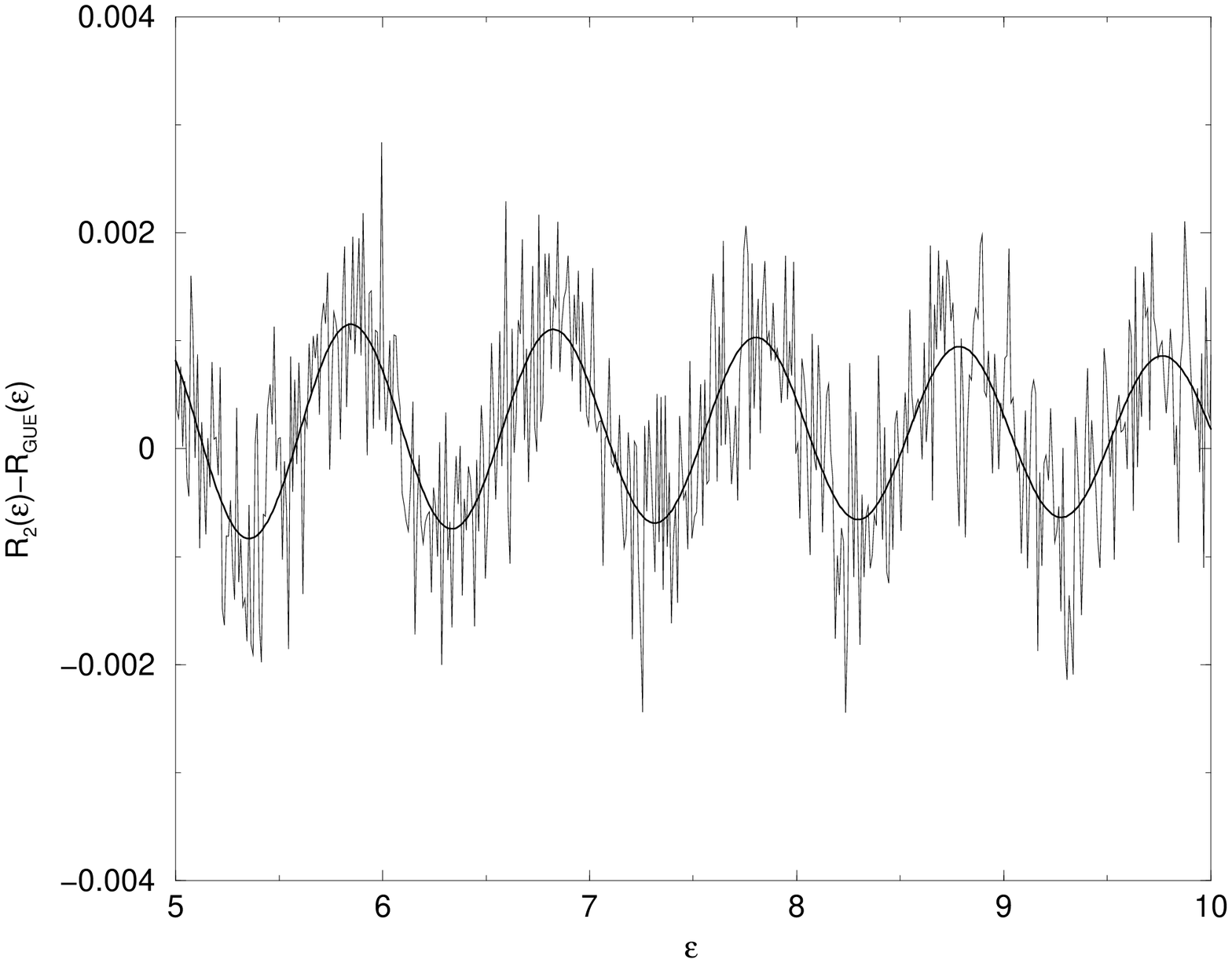}
\end{minipage}
\caption{The difference between the two-point  function of the
 Riemann zeros as in Fig.~\ref{fig2} and the GUE result for $0<\epsilon<5$ (left) and $5<\epsilon<10$ (right). The solid lines indicate  theoretical  predictions. }
 \label{fig5}
 \end{figure}

\begin{figure}
\begin{minipage}{.45\linewidth}
\includegraphics[angle=-90, width=.99\linewidth]{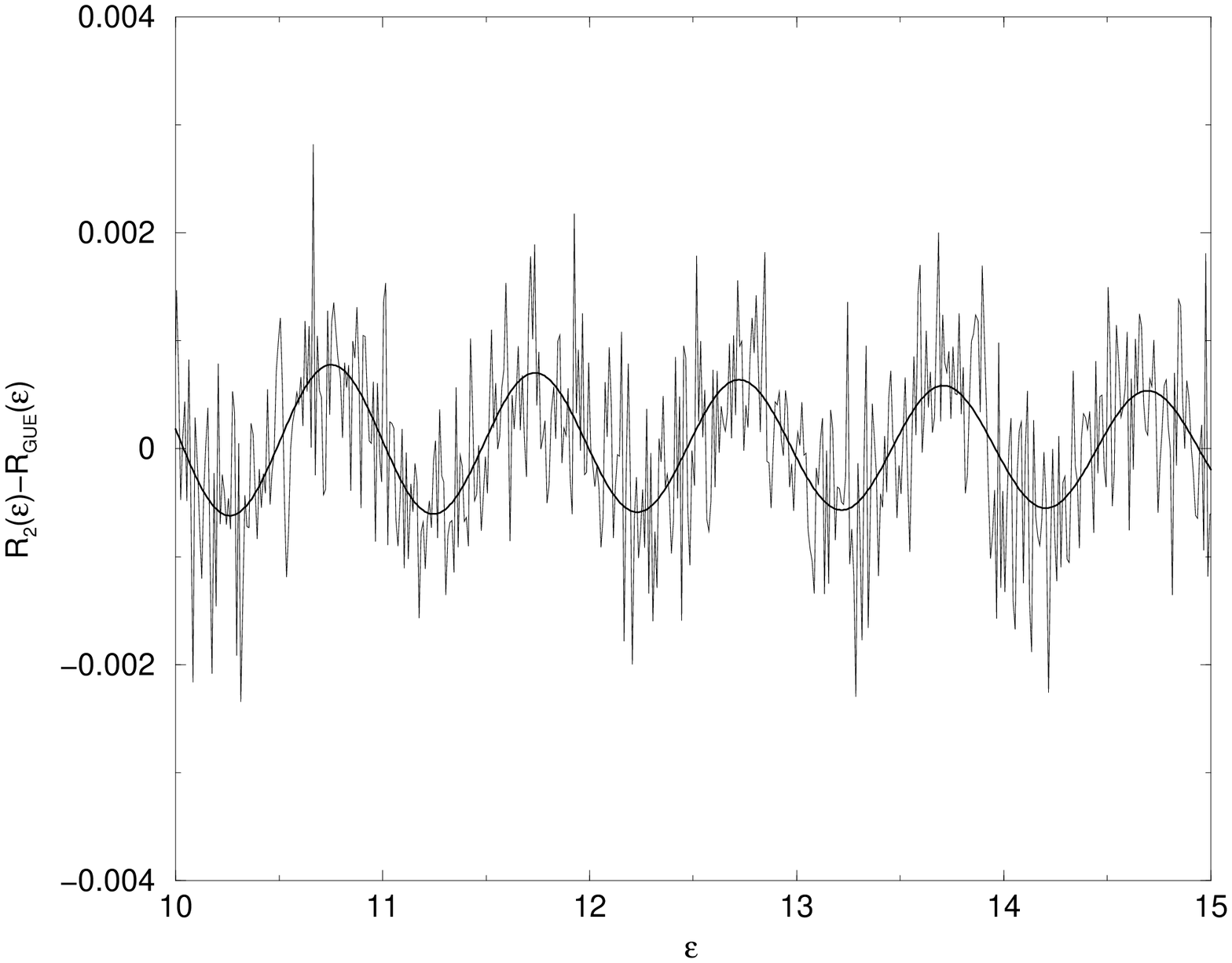}
\end{minipage}
\begin{minipage}{.45\linewidth}
\includegraphics[angle=-90, width=.99\linewidth]{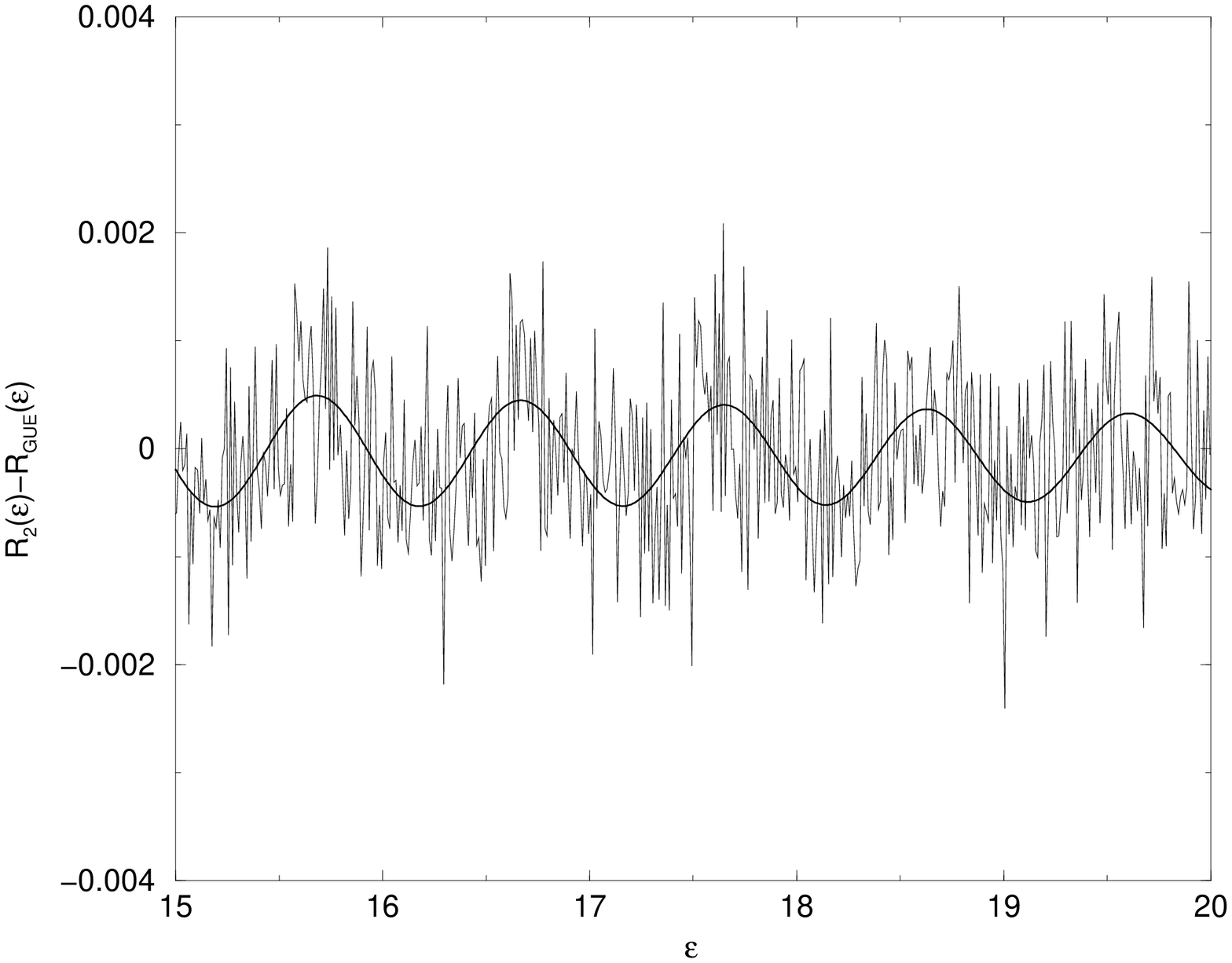}
\end{minipage}
\caption{The same as in Fig.~\ref{fig5} but in  the interval $10<\epsilon<15$ (left) and $15<\epsilon<20$ (right).}
\label{fig6}
\end{figure}

\begin{figure}[h]
\begin{minipage}{.45\linewidth}
\includegraphics[angle=-90, width=.99\linewidth]{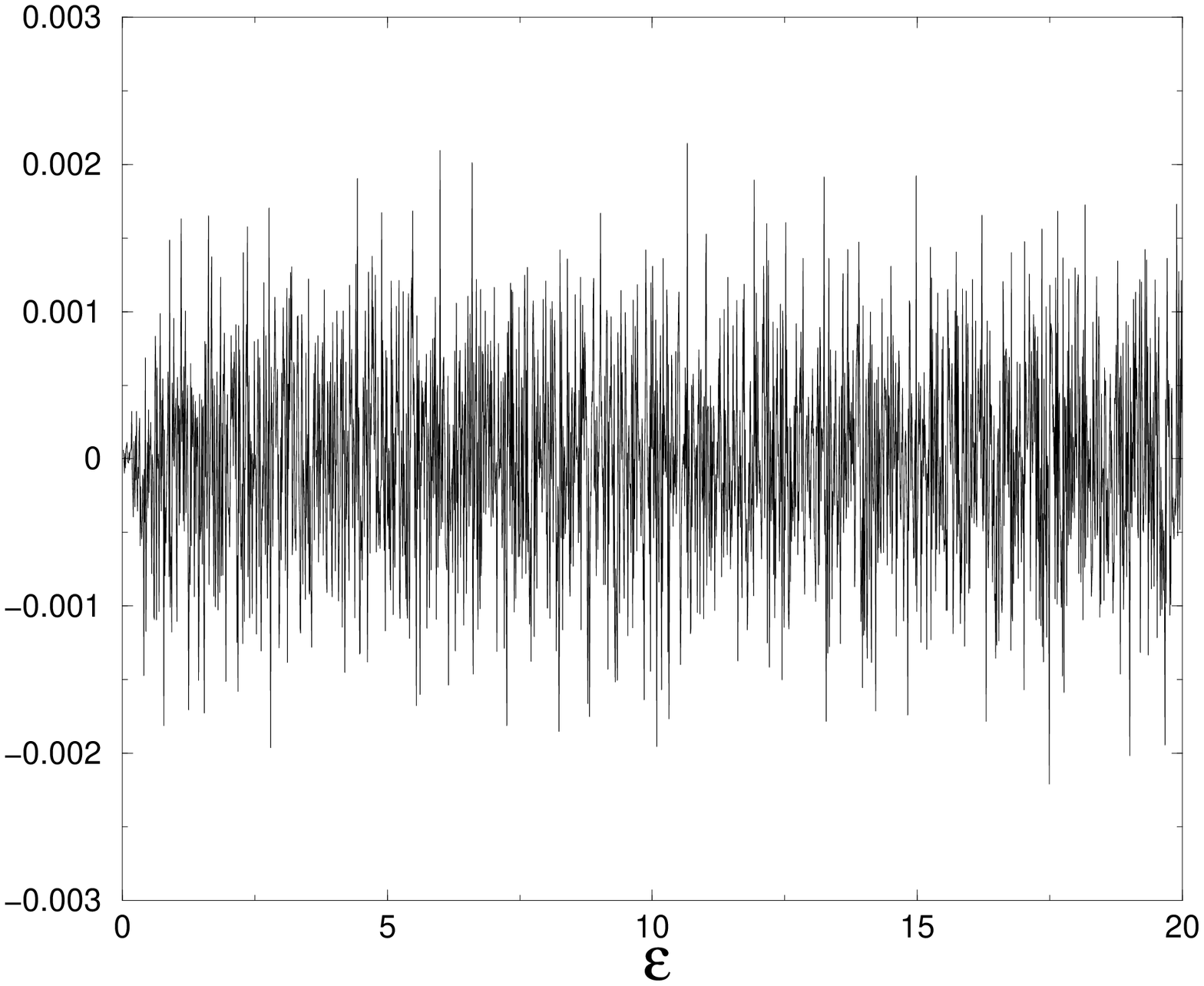}
\end{minipage}
\begin{minipage}{.45\linewidth}
\includegraphics[angle=-90, width=.99\linewidth]{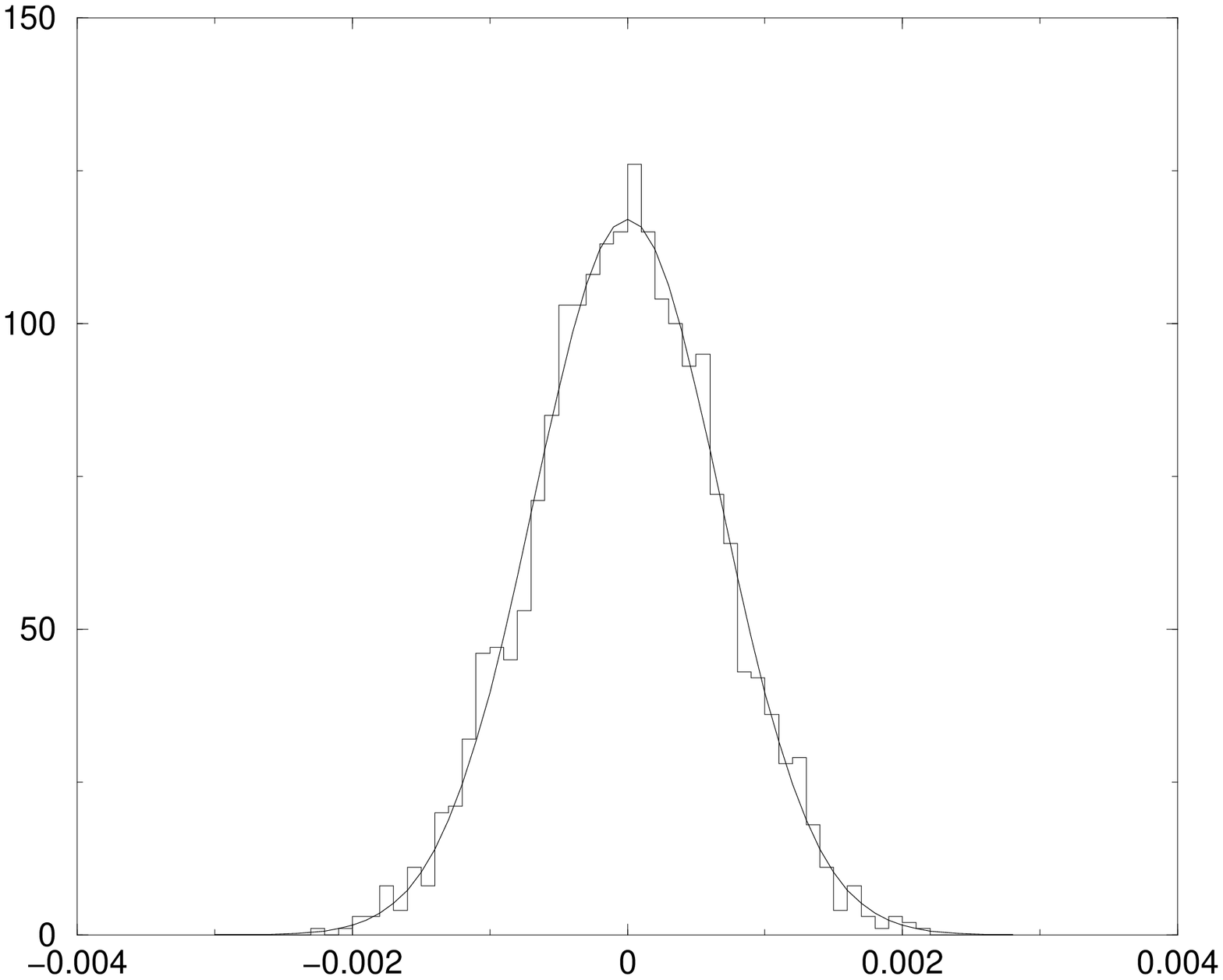}
\end{minipage}
\caption{Left: the difference between the two-point function computed by Odlyzko and the theoretical prediction  (\ref{exact_R}). Right:  The histogram of the previous differences. Solid line is the Gaussian fit.}
\label{fig7}
 \end{figure}

Taking into account the first corrections  to the GUE result one gets\cite{nearest_neighbor_2006}
\begin{equation}
R_2 (\varepsilon) = 1 - \frac{\sin^2 (\pi \varepsilon) }{ \pi^2 \varepsilon^2 } - 
\frac{\beta}{\pi^2 \bar{d}^2} \sin^2 (\pi \varepsilon) - \frac{\delta}{2 \pi^2
\bar{d}^3} \varepsilon \sin (2 \pi \varepsilon) + {\cal O} ( \bar{d}^{-4} )\; .
\label{expansion}
\end{equation}
Here $\beta$ and $\delta$ are numerical constants
\begin{equation} 
\beta=\gamma_0^2 + 2 \gamma_1 + \sum_p\frac{\ln^2 p}{(p-1)^2}\approx 1.57314\;,\;\;
\delta=\sum_p\frac{\ln^3 p}{(p-1)^2}\approx 2.3157
\label{constants}
\end{equation}
where the summation is performed over all prime numbers and  $\gamma_n$ are the Stieljes constants defined by  the  limit
\begin{equation}
 \gamma_n=\lim_{m\to \infty}\left (\sum_{k=1}^m\frac{\ln^n k}{k}-
\frac{\ln^{n+1} m}{m+1} \right )\;.
\end{equation} 
 From Fig.~\ref{fig2} it seems that  numerical calculations agree well with the GUE prediction. Nevertheless,  if one subtract the GUE result (\ref{R_2}) from these data it becomes evident  that in the difference there is a clear structure (see  Figs.~\ref{fig5} and \ref{fig6}). 
In Fig.~\ref{fig7} (left) we present the difference between numerical two-pooint correlation function and the derived formula for it (\ref{exact_R}). The result is structureless and its histogram is given in Fig.~\ref{fig7} (right). Solid line in this figure represents the Guassian fit to the histogram. The width of this Gaussian is close to the value of the bin used by Odlyzko  in the computation of the two-point function.  
This comparison is the best verification that the theoretical formula  (\ref{exact_R}) extremely well described monumental calculations of Odlyzko. The difference is purely statistical and shows no visible structure.   
 
 \section{Nearest-neighbor distribution}\label{nearest_neighbor}
 
There exist heuristic methods which permit the calculation of higher order correlation functions\cite{Varenna} but the resulting expressions become more and more tedious. For example, the  three-point correlation function of Riemann zeros has the form \cite{nearest_neighbor_2006}
\begin{equation}
R_3(e_1,e_2,e_3)=R_{3}^{({\rm diag})}(e_1,e_2,e_3)+R_3^{({\rm off})}(e_1,e_2,e_3)
\end{equation}
where the diagonal part is
\begin{eqnarray}
&&R_{3}^{({\rm diag})}(e_1,e_2,e_3)= -\frac{1}{(2\pi)^3}
\sum_p \log^3  p\left (
\frac{1}{(p^{1-{\rm i} e_{12}}-1)(p^{1-{\rm i} e_{13}}-1)}+\right .\nonumber\\
&&+  \left . 
\frac{1}{(p^{1-{\rm i} e_{21}}-1)(p^{1-{\rm i} e_{23}}-1)}+
\frac{1}{(p^{1-{\rm i }e_{32}}-1)(p^{1-{\rm i }e_{31}}-1)}\right )
+\mbox{c.c.} \;,
\end{eqnarray}
and the oscillating part is given by the following formula
\begin{eqnarray}
&&r_3^{({\rm off})}(e_1,e_2,e_3)=-\frac{1}{(2\pi {\rm  i})^3}
e^{2\pi {\rm i}\bar{d}e_{12}}|\zeta(1+{\rm i}e_{12})|^2
\prod_p\left (1 -\frac{(1-p^{{\rm i}e_{12}})^2}{(p-1)^2}\right )   \times \nonumber\\
&&\times \left [
\frac{\partial}{\partial e_3}\log 
\left |\frac{\zeta(1+{\rm i}e_{32})}{\zeta(1+{\rm i}e_{31})}\right |^2
-{\rm i} \sum_p \log p \left ( \frac{ p^{i e_{12}}-1}
{(p^{1+{\rm i} e_{23}}-1)(p^{1+{\rm i} e_{13}}-1)}\right .\right .+\nonumber\\
&&+\frac{p^{{\rm i} e_{12}}-1}{(p^{1-{\rm i} e_{13}}-1)(p^{1-{\rm i} e_{22}}-1)}+ \left . \left .\frac{(1-p^{{\rm i} e_{12}})^2}{p-2+p^{{\rm i} e_{12}}} 
(\frac{1}{p^{1-{\rm i}e_{31}}-1}+\frac{1}{p^{1-{\rm i}e_{23}}-1}) \right )\right ]\nonumber\\
&&+ \mbox{permutations}+\mbox{c.c.}\;.
\end{eqnarray}
Here $e_{ij}=e_i-e_j$ and it is assumed that the cyclic permutations of indices $1,2,3$ is taken together with the complex conjugation of the answer.
 
 The direct verification of these formulae is not easy but they can be used to obtain corrections to the other important statistics, namely, the nearest-neighbor distribution for Riemann zeros (cf. Fig.~\ref{fig2} (right)).  This can be done as follows\cite{nearest_neighbor_2006}.

For standard ensembles of random matrices correlation functions have the determinantal form (\ref{R_n}). In many cases from physical considerations it follows that first order  corrections to standard random matrix ensembles correspond to a change of the kernel $K(E_i,E_j)$  (\ref{kernel}) only.  In   Ref.~\citen{nearest_neighbor_2006} this statement has been checked using the two-point and the three-point correlation functions presented above and it was conjectured that it remains true for higher correlation functions of Riemann zeros as well.

From (\ref{expansion}) it is easy to check\cite{nearest_neighbor_2006} that the modified kernel takes the form
\begin{equation}
K(\varepsilon)=K_0(\varepsilon)+k_1(\varepsilon)
\end{equation}
where $K_0(\varepsilon)$ is the universal kernel (\ref{kernel}) and   $k_1(\varepsilon)$ is the effective correction to the universal result
\begin{equation}
k_1(\varepsilon)=\varepsilon\frac{\beta}{2\pi \bar{d}^2(E)} \sin(\pi \varepsilon)  +\varepsilon^2\frac{\delta}{2\pi
  \bar{d}^3(E)} \cos(\pi \varepsilon) = \frac{\pi \varepsilon}{6N_{\rm eff}^2} \sin(\pi \alpha \varepsilon)
\label{deltak}
\end{equation}
where  parameters $N_{\rm eff}$ and $\alpha$ have the following values
\begin{equation}
N_{\rm eff} = \frac{\pi \bar{d}(E) }{\sqrt{3 \beta} } \approx  \frac{\ln (E / 2
\pi) }{\sqrt{12 \beta} }\;,\;\;
\alpha = 1 + \frac{\delta}{2 \pi \bar{d}(E) \beta} =
1+\frac{\delta}{\beta \ln (E/2\pi)} \ .
\end{equation}
Here $\beta$ and $\delta$ are the same constants as in (\ref{constants}).

\begin{figure}
\includegraphics[width=.99\linewidth]{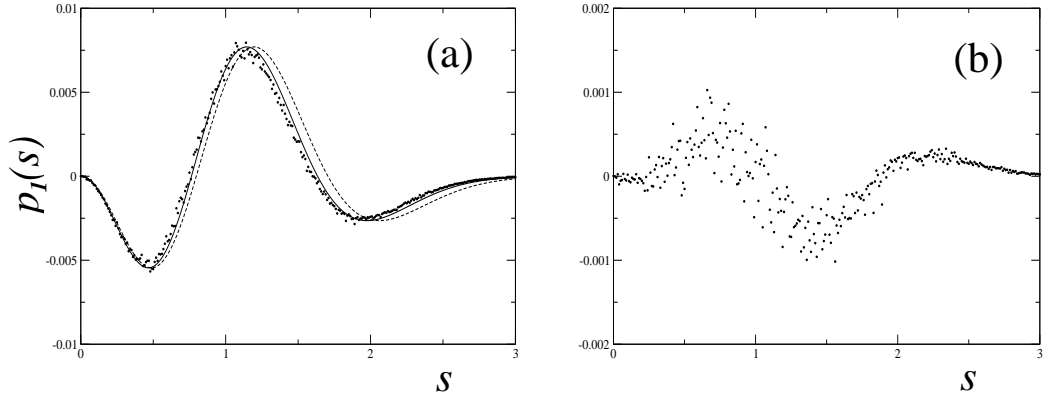}
\caption{{\small (a) Difference between the nearest neighbor spacing distribution
of the Riemann zeros and the asymptotic GUE distribution for a billion zeros
located in a window near $E=2.504 \times 10^{15}$ (dots), compared to the
theoretical prediction Eq.~(\ref{ps_correction}) (full line). The dashed line does not
include the scaling of $s$. (b) Difference between the numerical Riemann values
(dots) and the full curve (theory) of part (a).}}
\label{fig_ps_1}
\end{figure}

Using these expressions it was shown in Ref.~\citen{nearest_neighbor_2006}  that  the nearest--neighbor spacing distribution of the Riemann zeros can be calculated as follows. 
\begin{itemize}
\item 
First it is necessary to
find the expansion into powers of $N$ of the nearest--neighbor distribution for a circular unitary ensemble of $N\times N$ random unitary matrices equipped with the Haar measure (called CUE$_N$)  
\begin{equation}\label{pscue}
p^{\scriptscriptstyle {\rm (CUE_N)}} (s) = p_0 (s) + \frac{1}{N^2} \
p_1^{\scriptscriptstyle {\rm (CUE)}} (s) + {\cal O} ( N^{-4} ) \ ,
\end{equation}
\item 
Then the nearest--neighbor distribution for the Riemann zeros equals to the
universal random matrix result, $p_0(s)$ plus  the correction terms $\delta
p(s)$ where  
\begin{equation}
\delta p (s) = \frac{1}{N_{\rm eff}^2} \ p_1^{\scriptscriptstyle {\rm
(CUE)}} (\alpha s) + {\cal O} ( N_{\rm eff}^{-4} )\ .
\label{ps_correction} 
\end{equation}
with $N_{\rm eff}$ and $\alpha$ as above.
\end{itemize}
Fig.~\ref{fig_ps_1}(a) shows the comparison between  the difference of the numerical result of Odlyzko and  asymptotic formula 
Eq.~(\ref{ps_correction}) for a billion zeros located on a window around
 $E=2.504 \times 10^{15}$ (as in Fig.~\ref{fig2} right). The effective matrix size is $N_{\rm eff} = 7.7376$ and $\alpha = 1.0438$. The agreement is quite good.  For comparison, we have plotted as a dashed curve the theoretical formula (\ref{ps_correction}) without the rescaling of the $s$ variable. 

\begin{figure}
 \includegraphics[width=.99\linewidth]{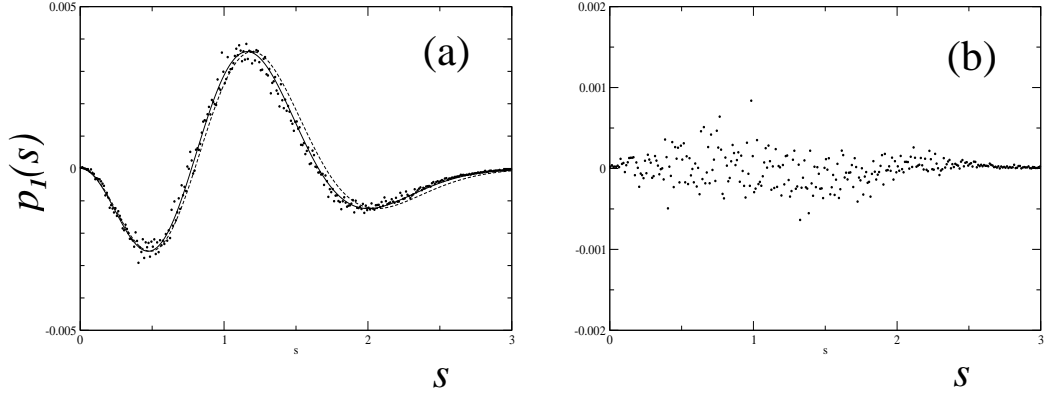}
 \caption{{\small Same as in Fig.~2 but for a billion zeros 
located in a window
near $E=1.307 \times 10^{22}$.}}
\label{fig_ps_2} 
\end{figure}

Fig.~\ref{fig_ps_1}(b)  is a plot of  the difference between Odlyzko's results and the prediction (\ref{ps_correction}).  There is still some structure visible 
which might be attributed to the ${\cal O} ( N_{\rm eff}^{-4} )$ correction. To test the
convergence, we have made the same plot but now using one billion zeros
located on a window around $E=1.307 \times 10^{22}$ which corresponds to 
$N_{\rm eff} = 11.2976$ and $\alpha = 1.0300$ (Fig.~\ref{fig_ps_2}(a)). Now the agreement is clearly  improved. The difference between the prediction (\ref{ps_correction}) and the numerical results, plotted in Fig.~\ref{fig_ps_2}(b) shows a structureless remain.
  
\section{ Moments of the Riemann zeta function}\label{zeta_moments}

The purpose of this Section is to discuss  the behavior of mean moments of the Riemann zeta function 
\begin{equation}
M_{\lambda}(T)=\frac{1}{T}\int_0^{T}|\zeta(1/2+{\rm i}t)|^{2\lambda}{\rm d}t
 \end{equation}
 when $T\to\infty$.

Naive calculations of this quantity can be performed as follows. Using the representation of $\zeta(s)$ as the Euler product (\ref{zeta_product}) one formally gets
\begin{equation}
|\zeta(1/2+{\rm i}t)|^{2\lambda}=\prod_p(1-A_p{\rm e}^{-{\rm i}E\ln p})^{-\lambda}(1-A_p{\rm e}^{{\rm i}E\ln p})^{-\lambda}
\label{moments}
\end{equation}
where $A_p=p^{-1/2}$. 

Logarithms of prime numbers, $\ln p$, with different primes are non-commensurated. Therefore, for any finite number of primes   ${\rm e}^{{\rm i}E\ln p}$ plays the role of random phase ${\rm e}^{{\rm i}\phi}$ and   
\begin{eqnarray}
&&\lim_{T\to \infty} \frac{1}{T}\int_0^{T}F({\rm e}^{{\rm i}E\ln p_1}, {\rm e}^{{\rm i}E\ln p_2},\ldots {\rm e}^{{\rm i}E\ln p_k}){\rm d}t=\\
&&=\int_0^{2\pi}\int_0^{2\pi}\ldots \int_0^{2\pi}F({\rm e}^{{\rm i}\phi_1},{\rm e}^{{\rm i}\phi_2},\ldots,{\rm e}^{{\rm i}\phi_k}) \frac{{\rm d}\phi_1}{2\pi}\frac{{\rm d}\phi_2}{2\pi}\ldots \frac{{\rm d}\phi_k}{2\pi}\;.
\nonumber
\end{eqnarray}
In application to (\ref{moments}) it is necessary only to compute the integral over factors with the same $p$ which can easily be done by the using the binomial expansion 
\begin{equation}
(1+x)^{-\lambda}=\sum_{m=0}^{\infty} (-x)^m
\frac{\Gamma(m+\lambda)}{m!\Gamma(\lambda) }\;.
\end{equation}
The result is
\begin{equation}
\int_0^{2\pi}(1-A_p{\rm e}^{-{\rm i}\phi_p})^{-\lambda}(1-A_p{\rm e}^{{\rm i}\phi_p})^{-\lambda}\frac{{\rm d}\phi_p}{2\pi}=\sum_{m=0}^{\infty}
 A_p^{2m}\left (\frac{\Gamma(m+\lambda)}{m!\Gamma(\lambda)}\right )^2 .
 \label{factor}
\end{equation} 
Finally, it is tempting to conclude that mean moments of the Riemann zeta function are the product over all prime of factors (\ref{factor})
\begin{equation}
M_{\lambda}(T)\stackrel{T\to\infty}{\longrightarrow} \prod_p \sum_{m=0}^{\infty}
\left (\frac{\Gamma(m+\lambda)}{m!\Gamma(\lambda)}\right )^2p^{-m}\;.
\label{naive}
\end{equation}
Unfortunately, at the critical line (i.e. when $A_p=p^{-1/2}$) the product diverges and the divergent part is $\prod_p(1+\lambda^2/p)$. To get a finite expression let us define
\begin{equation}
a(\lambda)=\prod_p ((1-1/p)^{\lambda^2}\sum_{m=0}^{\infty}
\left (\frac{\Gamma(m+\lambda)}{m!\Gamma(\lambda)}\right )^2p^{-m}\;.
\end{equation}
The factor  $(1-1/p)^{\lambda^2}$ ensures the convergence of the infinite product. 

Due to the divergence of the product in (\ref{naive}) it was evident that certain increasing factors of $T$ have to be introduced. So it was conjectured that instead of (\ref{naive}) one should have the following result
\begin{equation} 
\lim_{T\to\infty} \frac{1}{(\ln T)^{\lambda^2}}M_{\lambda}(T)=a(\lambda)f(\lambda)
 \end{equation}
 with an unknown function $f(\lambda)$. 
 
 This conjecture has been based on the two proved theorems: $f(0)=1$ by definition,  $f(1)=1$ (Hardy and Littlewood, 1918), $f(2)=1/12$ (Ingram, 1926), and two conjectures: $f(3)=42/9! $ (Conrey, Ghosh, 1992) and $f(4)=24024/16!$ (Conrey and Gonek,  1998). 
 
 The breakthrough in this problem was the work of Keating and Snaith \cite{Keating_Snaith}.  Their argumentation was as follows.  By definition, 
 $\zeta(1/2+{\rm i}E_j)=0$ where $E_j$ are non-trivial zeros . Therefore, one can write $\zeta(1/2+{\rm i}E)\sim \prod_j(E-E_j)$. If such zeros are in a certain sense eigenvalues of an unitary matrix $U$ this representation means that the zeta function is proportional to a characteristic polynomial of such matrix i.e. $\zeta(1/2+{\rm i}E)\sim \det({\rm e}^{{\rm i}E}-U)$. But for random matrices from standard  ensembles the moments of  characteristic polynomials can be computed explicitly.
 
 In partricular for $N\times N$ unitary matrices equipped with the Haar measure the calculations are straighforward. By definition one gets
\begin{equation}
\left <|Z|^{2\lambda}\right >_{U(N)}= \frac{1}{N!}\left (\prod_{k=1}^N\int_0^{2\pi}
\frac{{\rm d}\theta_k}{2\pi}\right ) \prod_{1\leq j<m\leq N}|{\rm e}^{{\rm i}\theta_j}-{\rm e}^{{\rm i}\theta_m} |^2
\left |\prod_{n=1}^{N}({\rm e}^{{\rm i}\theta}-{\rm e}^{{\rm i}\theta_n})\right |^{2\lambda}.
\end{equation}
The integrals can be computed from the Selberg integral (see e.g. Ref.~\citen{Mehta}) and the result is the following
\begin{equation}
\left <|Z|^{2\lambda}\right >_{U(N)}=\prod_{j=1}^N
\frac{\Gamma(j)\Gamma(2\lambda+j)}{(\Gamma(j+\lambda))^2}\;.
\end{equation}
 When $N\to\infty$  the right-hand side of this expression tends to
\begin{equation}
\left <|Z|^{2\lambda}\right >_{U(N)}\stackrel{N\to \infty}{\longrightarrow} N^{\lambda^2}f_{COE}(\lambda)
\end{equation}
where the function $f_{COE}(\lambda)$ is 
\begin{equation}
f_{COE}(\lambda)=\frac{G^2(1+\lambda)}{G(1+2\lambda)}
\end{equation}
and $G(z)$ is the Barnes $G$-function
\begin{equation}
G(1+z)=(2\pi)^{z/2}{\rm e}^{-[(1+\gamma)z^2+z]/2}\prod_{n=1}^{\infty}\left [ (1+z/n)^n {\rm e}^{-z+z^2/(2n)}\right ]\;.
\end{equation}
By analogy with $\Gamma(z)$ function the function $G(z)$ can be defined as follows $G(1)=1$ and $G(z+1)=\Gamma(z)G(z)$. (Cf.  $\Gamma(1)=1$ and $\Gamma(z+1)=z\Gamma(z)$.)

Let us compute the first values of   $f_{COE}(\lambda)$: $f_{COE}(0)=1$, $f_{COE}(1)=1$, $f_{COE}(1)=1/12$, $f_{COE}(3)=42/9!$, $f_{COE}(4)=24024/16!$. Comparing with proved and conjectured values of $f(\lambda)$ cited above Keating and Snaith had conjectured that for general $\lambda$ 
\begin{equation}
f(\lambda)=f_{COE}(\lambda)
\end{equation}
and  the matrix dimension, $N$,  asymptotically is related with $T$ as 
\begin{equation}
N=\ln T\;.
\end{equation}
Under this identification all known and conjectured values of zeta function moments are in agreement with values obtained from the random matrix theory.
  
\section{ Conclusion}\label{conclusion}

\begin{itemize}
\item Quantum chaos methods and ideas  are extremely successively applied to the Riemann zeta function and other zeta functions and $L$-functions from number theory.
\item Obtained heuristic formulae are very well confirmed by existing numerics. Even tiny details of monumental numerical calculations of Odlyzko are explained by these formulae.
\item There exists a quite large number of random matrix type conjectures for 'all' possible quantities like spectral statistics, moments, ratios of different zeta functions etc.
\item These conjectures are well accepted but no one of such results is proved rigorously. 
\end{itemize}

\end{document}